\documentclass[prd,twocolumn,showpacs,nofootinbib,amsmath,amssymb,floatfix,superscriptaddress,showkeys]{revtex4-1}
\usepackage{multirow}
\usepackage{graphicx}
\usepackage{epstopdf}
\usepackage{dcolumn}
\usepackage{bm}
\usepackage{threeparttable}
\usepackage{subfigure}
\usepackage{color,txfonts}
\usepackage{ulem}
\definecolor{blue0}{rgb}{0,0,0.6}
\usepackage{multirow}
\usepackage[colorlinks,linkcolor=blue0,anchorcolor=blue0,citecolor=blue0,urlcolor=blue0]{hyperref}

\newcommand{\jcap}{J. Cosmol. Astropart. Phys.}
\newcommand{\physrep}{Phys. Rep.}
\newcommand{\mnras}{Mon. Not. R. Astron. Soc.}

\newcommand{\aap}{Astron. Astrophys}
\newcommand{\aj}{Astron. J.}

\newcommand{\pasj} {PASJ}

\newcommand{\apjl}{Astrophys. J.}

\begin{document}

\title{Constraints on the annihilation of heavy dark matter in dwarf spheroidal galaxies with gamma-ray observations}

\author{Xiao-Song Hu}
\author{Ben-Yang Zhu}
\author{Tian-Ci Liu}
\author{Yun-Feng Liang}
\email[]{liangyf@gxu.edu.cn}
\affiliation{Guangxi Key Laboratory for Relativistic Astrophysics, School of Physical Science and Technology, Guangxi University, Nanning 530004, China}

\date{\today}

\begin{abstract}
Electrons and positrons produced in dark matter annihilation can generate secondary emission through synchrotron and IC processes, and such secondary emission provides a possible means to detect DM particles with masses beyond the detector’s energy band.
The secondary emission of heavy dark matter (HDM) particles in the TeV-PeV mass range lies within the Fermi-LAT energy band. In this paper, we utilize the Fermi-LAT observations of dwarf spheroidal (dSph) galaxies to search for annihilation signals of HDM particles. We consider the propagation of $e^+/e^-$ produced by DM annihilation within the dSphs, derive the electron spectrum of the equilibrium state by solving the propagation equation, and then compute the gamma-ray signals produced by the $e^+/e^-$ population through the IC and synchrotron processes.
{Considering the spatial diffusion of electrons, the dSphs are modeled as extended sources in the analysis of Fermi-LAT data according to the expected spatial intensity distribution of the gamma rays.}
We do not detect any significant HDM signal. By assuming a magnetic field strength of $B=1\,{\rm \mu G}$ and a diffusion coefficient of $D_0 = 3\times10^{28}\,{\rm cm^{2}s^{-1}}$ of the dSphs, we place limits on the annihilation cross section for HDM particles. 
{Our results are weaker than the previous limits given by the VERITAS and IceCube observations of dSphs, but extend the existing limits to higher DM masses.} As a complement, we also search for the prompt $\gamma$-rays produced by DM annihilation and give limits on the cross section in the 10-$10^5$ GeV mass range. Consequently, in this paper we obtain the upper limits on the DM annihilation cross section for a very wide mass range from 10 GeV to 100 PeV in a unified framework of the Fermi-LAT data analysis.
\end{abstract}
\maketitle

\section{introduction}
Dark matter (DM) is one of the most intriguing puzzles in modern physics. It constitutes about 27\% of the total energy density of the Universe \cite{Planck2015}, but its nature remains elusive. Various astrophysical observations, such as the cosmic microwave background power spectrum, the rotation curves of galaxies, and the gravitational lensing of galaxy clusters, indicate that DM is nonbaryonic and cold (cold means that it has a negligible thermal velocity). A promising class of DM candidates is weakly interacting massive particles (WIMPs), which can self-annihilate or decay into standard model (SM) particles, producing $\gamma$ rays and cosmic rays {(CRs)} that can be detected by space-based instruments \cite{SupersymmetricDM,ParticleDM}. Some examples of these instruments are the Fermi Large Area Telescope (Fermi-LAT \cite{FermiLat,Fermi2015prl}), the Dark Matter Particle Explorer (DAMPE \cite{Dampe,FAN201883}), the Alpha Magnetic Spectrometer (AMS-02 \cite{AMS02HEAP,AMS02pm}).

The kinematic observations show that dwarf spheroidal (dSph) galaxies are DM-dominant systems. DSphs are ideal targets for indirect DM searches, as they have low astrophysical backgrounds and lack conventional $\gamma$-ray production mechanisms, unlike the galactic center where the DM signal is obscured by large uncertainties of the diffuse emission and complex astrophysical backgrounds \cite{1990Natur.346...39L,2013PhR...531....1S,PhysRevD.69.123501}. Currently, more than 60 dSphs or candidates have been discovered by wide-field optical imaging surveys\cite{2000AJ....120.1579Y,Belokurov_2007,2018A&A...616A...1G,2015ApJ...807...50B}
Many groups have used Fermi-LAT data to search for $\gamma$-ray emission from dSphs with different methods and assumptions, but no significant signals have been found so far, leading to stringent constraints on the mass $m_{\chi}$ and the annihilation cross section $\langle\sigma v\rangle$ of DM particles \cite{2011PhRvL.107x1302A,2011PhRvL.107x1303G,2012PhRvD..86b3528C,2013JCAP...03..018S,2014PhRvD..89d2001A,Fermi2015prl,2015PhRvD..91h3535G,2016PhRvD..93h3513Z,2020JCAP...02..012H,Zhu:2022scj,PhysRevD.106.123032}. 

Alternatively, one can also search for DM annihilation signals in the form of synchrotron and inverse Compton (IC) emission from the cosmic-ray electrons and positrons produced by the DM {annihilation \cite{2006A&A...455...21C,2007PhRvD..75b3513C,2020PhRvD.101b3015K,2023arXiv230314117W,2017JCAP...09..027M,Vollmann_2021,2021JCAP...09..025C,PhysRevD.107.103011,2023JCAP...08..030R,2014JCAP...10..016R,2017JCAP...07..025R}.}
DM annihilation can produce various SM particles, such as quarks, leptons, and bosons, which can further decay or hadronize into electrons and positrons \cite{ParticleDM}. These charged particles can radiate synchrotron photons in the presence of magnetic fields and also scatter with the ambient photons to produce IC photons. This method can potentially probe DM in a wider mass range than the direct $\gamma$-ray emission, as the secondary radiation spectrum peaks at energies different from the prompt annihilation emission. However, this method also suffers from some astrophysical uncertainties, such as the magnetic field strength and distribution, and the diffusion coefficient of the {CR}, which need careful modeling.

Heavy dark matter (HDM) has a DM mass ranging from 10 TeV to the Planck energy ($\sim$ $10^{19}$ GeV). In recent years, HDM has attracted a lot of attention and many researchers have searched for HDM using different techniques and data sets \cite{2016PhRvD..94f3535K,aartsen2018search,2017PhRvL.119b1102C,2022PhRvL.129z1103C,2022arXiv220512950A,2023ApJ...945..101A,2023PhRvD.107l3027Z,PhysRevD.108.043001}. But for the annihilation HDM, the particle mass of thermal relic dark matter is constrained by the unitarity bound, which is derived from the thermal production mechanism and the quantum principle of probability conservation \cite{1990PhRvL..64..615G,2022ApJ...938L...4T}. This bound has discouraged people from exploring annihilation HDM, but there are several mechanisms have been proposed to violate this bound \cite{2022arXiv220306508C}. For the $e^+/e^-$ cosmic-ray particles from DM annihilation, their secondary emission produced by TeV-PeV HDM falls within the Fermi-LAT sensitivity range, which makes it possible to probe HDM with Fermi-LAT data. 

In this work, we use 14 years of Fermi-LAT observation data to constrain the parameter space of the HDM. We analyze the secondary synchrotron and IC emission from 8 classical dSphs that have more reliable DM halo parameters. 
We take into account both the astrophysical parameters and the $e^+/e^-$ spectrum of HDM annihilation, and perform the analysis with relatively conservative assumptions. 
As a complement, we also search for the prompt $\gamma$-ray signals of DM annihilation from a larger sample of 15 dSphs.  
Using the secondary emission, we present the 95\% confidence level (CL) constraints on the cross section in a mass range of $100$ TeV - $10^5$ TeV. 
Based on the prompt $\gamma$-ray emission, we derive the upper limits on the annihilation cross section for the $b\bar{b}$ and $\tau^{+}\tau^{-}$ channels for DM masses from 10 GeV to $10^5$ GeV. To enhance the sensitivity, we perform a combined likelihood analysis.

\section{Synchrotron and IC Radiation from Dark Matter Annihilation}
The synchrotron and IC emission from DM annihilation have been extensively studied \cite{2006A&A...455...21C,2007PhRvD..75b3513C,2020PhRvD.101b3015K,2023arXiv230314117W,2017JCAP...09..027M,Vollmann_2021,2021JCAP...09..025C,PhysRevD.107.103011,2023JCAP...08..030R,2014JCAP...10..016R,2017JCAP...07..025R}. When a DM pair annihilates into SM particles ($b\bar{b},\tau^{+}\tau^{-},t\bar{t}$, etc) in a dSph, these particles can further decay or hadronize into electrons and positrons. The electrons and positrons then diffuse through the interstellar medium in the galaxy and lose energy through various processes. To calculate the synchrotron and IC emission, we need to obtain the equilibrium spectrum of the electrons and positrons. Assuming steady-state and homogeneous diffusion, the propagation equation can be written as\footnote{{We neglect the advection and re-acceleration effects which are subdominant at the energies we are interested in. Please see Appendix~\ref{app:b}.}} \cite{2017JCAP...09..027M,2020PhRvD.101b3015K}
\begin{equation}
D(E)\nabla^{2}\left(\frac{d n}{d E}\right)+\frac{\partial}{\partial E}\left(b(E,r)\frac{d n}{d E}\right)+Q_{e}(E,r)=0
\label{propagationequation}
\end{equation}
where $n$ is the equilibrium electron density, $D(E)$ is the diffusion coefficient, $b(E)$ is the energy loss term and $Q_{e}(E,r)$ is the electron injection from DM annihilation. The injection term $Q_{e}(E,r)$ of DM annihilation is
\begin{equation}
	Q_{e}(E,r)=\frac{\langle\sigma v\rangle}{2}\frac{\rho(r)^2}{m_{\chi}^2}\frac{d N_e}{d E}
	\label{injection}
\end{equation}
where ${dN_e}/{dE}$ is the $e^{+}/e^{-}$ yield per DM annihilation, which can be obtained from {\tt HDMSpectra} \cite{2021JHEP...06..121B}, a Python package provides HDM spectra from the electroweak to the Planck scale. The $\rho(r)$ is the DM density profile of the dSph. N-body cosmological simulations suggest that it can be described by a Navarro-Frenk-White (NFW) profile \cite{1997ApJ...490..493N}. For the DM density distribution, we adopt the NFW profile,
\begin{equation}
	\rho_{\mathrm{NFW}}(r)=\frac{\rho_{s}r_{s}^{3}}{r(r_{s}+r)^{2}}.
	\label{nfwprofile}
\end{equation}
The $r_s$ and $\rho_s$ for the 8 classical dSphs are listed in Table~\ref{TableI}. 
For the diffusion term $D(E)$, we assume a power-law dependence on the energy as follows,
\begin{equation}
D(E,r)=D_0\left({\frac{E}{\rm{GeV}}}\right)^{\delta},
\end{equation}
where $D_0$ is the diffusion coefficient and $\delta$ is the index. In the Milky Way, {CRs} are generally assumed to have $D_0 \sim 10^{28}$ cm$^{2}$s$^{-1}$ and $\delta$ in a range of 0.3-0.7 \cite{2020A&A...639A.131W,2019PhLB..789..292W,2019PhRvD..99j3023E,1971ApJ...163..255P,2016ApJ...824...16J}. For dSphs, the situation may be more complex. In this work, we use $D_0 = 3\times10^{28}$ cm$^{2}$s$^{-1}$ and $\delta = 0.3$, which are relatively moderate values \cite{2020PhRvD.101b3015K}. {The dependence of our results on the value of $D_0$ will be discussed in Sec.~\ref{sec:res}.}

The main processes for the energy loss of $e^+/e^-$ include synchrotron radiation, IC scattering, Coulomb scattering, and bremsstrahlung. The 
energy loss rate $b(E)$ can be described as \cite{2017JCAP...09..027M,2020PhRvD.101b3015K}:
\begin{equation}
\begin{aligned}
b(E,r) &=b_{\rm{IC}}(E)+b_{\rm{syn}}(E,\mathbf{r})+b_{\rm{Coul}}(E)+b_{\rm{brem}}(E) \\& =b_{\rm{IC}}(E) +b^0_{\mathrm{\rm{syn}}}\left(\frac{E}{\rm GeV}\right)^2\left(\frac{B}{\mu G}\right)^2 \\& +b^0_{\rm{Coul}} n_{e}\left[1+\log\left(\frac{E/m_e}{n_e}\right)/75\right] \\& + b^0_{\rm{brem}}n_e\left[\log\left(\frac{E/m_e}{n_e}\right)+0.36\right]
\label{energylossterm}
\end{aligned}
\end{equation}
where $m_e$ is the electron mass, $n_e$ is the mean number density of thermal electrons and it is about $10^{-6}$ cm$^{-3}$ in dSphs \cite{2015MNRAS.448.3747R}. 
{The energy loss coefficients are taken to be $b^0_{\rm{syn}} \simeq 0.0254$, $b^0_{\rm{Coul}} \simeq 6.13$, $b^0_{\rm{brem}} \simeq 1.51$, all in units of $10^{-16}$ GeVs$^{-1}$ \cite{2006A&A...455...21C,2007PhRvD..75b3513C}}. In fact, in the energy range we are considering, only the IC and synchrotron processes are important. 

{For the IC energy loss, the Klein–Nishina (KN) effect is important at high energies, the IC loss rate can be expressed as \cite{2010NJPh...12c3044S}
\begin{equation}
	b_{\rm IC}(E) = \frac{20 \sigma_{T} c \rm W_{\rm CMB}}{\pi^4} \gamma^2 I(\gamma,T)
\end{equation}
where $\sigma_{T}$ is the Thomson cross section, $\gamma = {\rm{ E}}/(m_ec^2)$ is the Lorentz factor. We only consider the scattering on the CMB photons, which has a energy density of $\rm W_{ CMB} \approx 0.26 \ \rm eVcm^{-3}$, and 
\begin{equation}
	I(\gamma,T) = \frac{9}{(kT)^4} \int_{0}^{\infty} d\epsilon \frac{\epsilon^3}{{\rm {exp}}(\epsilon/kT)-1}\int_{0}^{1} \frac{q F(\Gamma,q)}{(1+\Gamma q)^3} 
\end{equation}
with
\begin{align}
    F(\Gamma,q) &= 2 q \ln{q} +(1+2q)(1-q) + \frac{(\Gamma q)^2 (1-q)}{2(1+\Gamma q)}, \\ \nonumber
    & \Gamma = 4 \epsilon \gamma/(m_e c^2),\ \ q = \epsilon/\Gamma (\gamma m_ec^2-\epsilon)
\end{align}
where $k$ is the Boltzmann constant, $T=2.725\,{\rm K}$ is the CMB temperature, $\epsilon$ is the energy of the target photon.}

The energy loss term of synchrotron depends on the galactic magnetic field. Previous research on the magnetic fields in dSph galaxies indicates that the strength is at a $\mu$G level \cite{2007PhRvD..75b3513C,2013ApJ...773...61S}. We also have little knowledge of the spatial profile of the magnetic fields in dSphs. In this work, we consider a uniform magnetic field within each dSph and adopt the value of $B = 1\,\mu$G for the field strength \cite{2007PhRvD..75b3513C}.
{The influence of $B$ on the results will be discussed in Sec.~\ref{sec:res}.}

\begin{table}[b]
	\centering
	\caption{Parameters of 8 Milky Way classical dSphs.}
	\begin{ruledtabular}
		\begin{tabular}{lccccc}
	       Name & \multicolumn{1}{c}{\begin{tabular}[c]{@{}c@{}}$r_{h}$ \\ (kpc)\end{tabular}} & \multicolumn{1}{c}{\begin{tabular}[c]{@{}c@{}}$r_{s}$\\(kpc)\end{tabular}}& \multicolumn{1}{c}{\begin{tabular}[c]{@{}c@{}}$\rho_{s}$ \\ (GeV/cm$^3$)\end{tabular}} & \multicolumn{1}{c}{\begin{tabular}[c]{@{}c@{}}$\theta_h$ \\ (deg)\end{tabular}} &\multicolumn{1}{c}{\begin{tabular}[c]{@{}c@{}}$\theta_{68}$ \\ (deg)\end{tabular}} \\
			\colrule
			Ursa Minor & 3.2 & 2.2 & 0.8 & 2.4 & 1.2 \\
			Sculptor & 5.3 & 2.1 & 0.9 & 3.5 & 1.5 \\
			Sextans & 5.1 & 1.1 & 1.0 & 3.4 & 1.4\\
			Leo I & 3.9 & 1.5 & 0.9 & 0.9 & 0.4\\
			Leo II & 1.6 & 1.0 & 1.6 & 0.4 & 0.2\\
			  Carina & 4.5 & 1.7 & 0.6 & 2.5 & 1.1\\
			Fornax & 12.5 & 2.8 & 0.5 & 4.9 & 1.0\\
			Draco & 2.5 & 1.0 & 1.4 & 1.9 & 1.0 \\
		\end{tabular}
	\end{ruledtabular}
\begin{tablenotes}
	\footnotesize
	\item Note: 
	The values of $r_h$, $r_{s}$ and $\rho_{s}$ for the dSphs except Draco are adopted from \cite{Geringer-Sameth_2015,PhysRevD.106.123032}. For the Draco dSph, we use the data from \cite{2007PhRvD..75b3513C}. 
 The coordinates and distance of Leo I are $(l=226.0^\circ, b=49.1^\circ)$ and 254 kpc, respectively.
 Such information for the other 7 dSphs can be found in Table~\ref{TableII}.
 {Due to the diffusion of CR electrons, the gamma-ray emission from DM in each dSph appears as an extended source in the sky.
The $\theta_{68}$ is the 68\% containment angle of the extended gamma-ray emission at 500 MeV under benchmark parameters ($D_0=3\times10^{28}\,\rm cm^2\,s^{-1}$, $B=1\,\rm \mu G$). The $\theta_h$ is the angular radius corresponding to $r_h$.}
\end{tablenotes}
\label{TableI}
\end{table}

\begin{figure*}[htbp]
	\centering
	\includegraphics[width=0.45\textwidth]{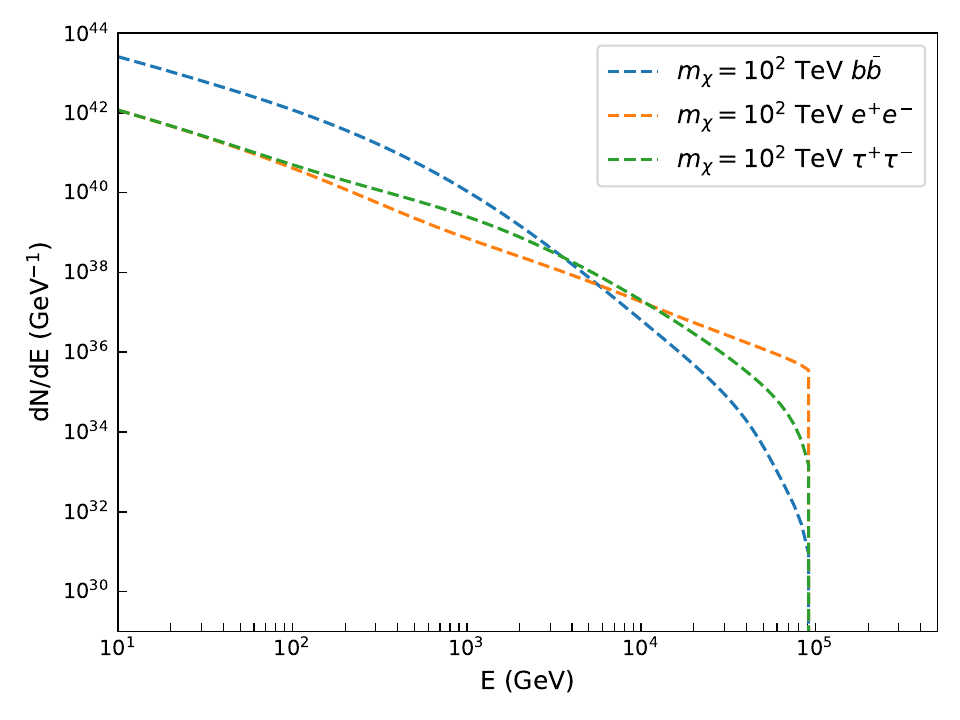}
	\includegraphics[width=0.45\textwidth]{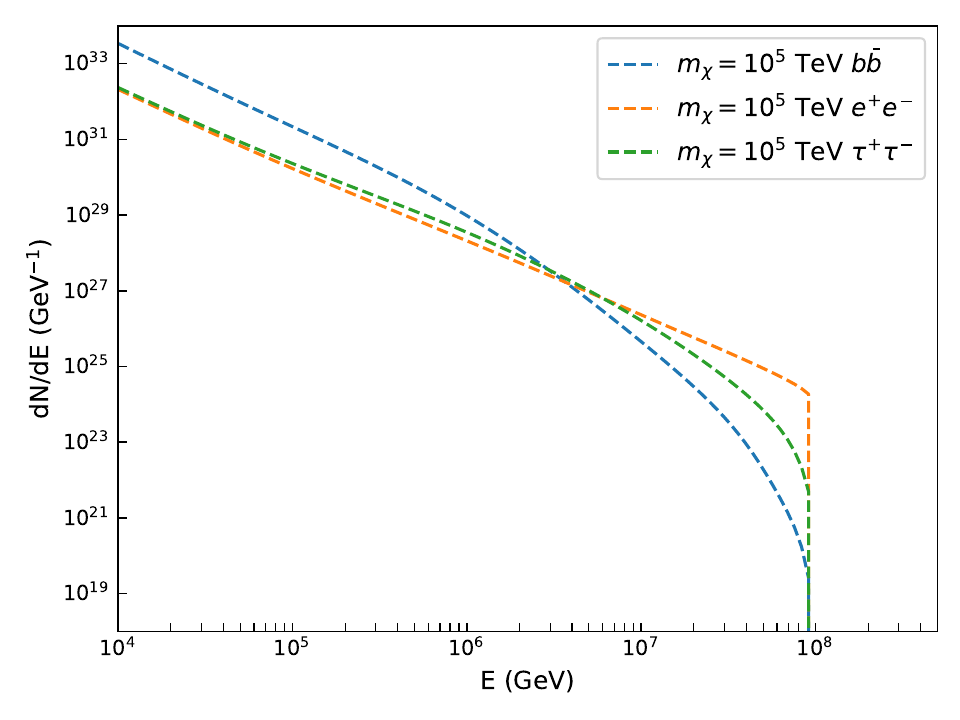}
	\caption{The electron spectra as a function of electron energy from the DM annihilation in Draco dSph for different annihilation channels. Left panel: $m_\chi$ = 10$^2$ TeV. Right panel: $m_\chi$ = 10$^5$ TeV. Three annihilation channels are shown: $b\bar{b}$ (blue line), $\tau^{+}\tau^{-}$ (green line) and $e^{+}e^{-}$ (orange line). The magnetic field strength is $B = 1\,{\rm \mu G}$, the diffusion coefficient is $D_0 = 3\times10^{28}\,{\rm cm^{2}s^{-1}}$ and the annihilation cross section is $\langle \sigma v\rangle = 10^{-26}$ cm$^{3}$s$^{-1}$.}
	\label{fig:e_distribution}
\end{figure*}

With the boundary condition of $dn/dE(r_h)=0$, an analytic solution for Eq.~(\ref{propagationequation}) has been obtained using the Green function method \cite{2007PhRvD..75b3513C}. For the dSphs, we consider the steady-state case. The solution of Eq. (\ref{propagationequation}) is given by,
\begin{equation}
\frac{d n}{d E}(r,E)=\frac{1}{b(E)}\int_{E}^{m_{\chi}}d E^{\prime}G(r,\Delta v)Q_{e}(E^{\prime},r).
\label{electrondnde}
\end{equation}
The Green function $G(r,\Delta v)$, is obtained by using the method of image charges. More details of the derivation of the Green function can be found in Ref.~\cite{2006A&A...455...21C}. A new method that does not rely on the image charge technique, but instead uses a Fourier series expansion of the Green function, has been proposed by Ref. \cite{Vollmann_2021}. Some groups have verified that the maximum difference between the two numerical methods is around 40\% $\sim$ 50\% \cite{Bhattacharjee_2021,2021JCAP...09..025C}. The free-space Green function can be expressed as \cite{2017JCAP...09..027M,2020PhRvD.101b3015K},
\begin{equation}
\begin{aligned}
G(r,\Delta v) &=\frac{1}{\sqrt{4\pi\Delta v}}\sum_{n=-\infty}^{n=\infty}(-1)^{n}\int_{0}^{r_{h}}dr^{\prime}\frac{r^{\prime}}{r_{n}}\left(\frac{\rho(r^{\prime})}{\rho(r)}\right)^{2} \\& \left[\exp\left(-\frac{\left(r^{\prime}-r_{n}\right)^{2}}{4\Delta v}\right)-\exp\left(-\frac{\left(r^{\prime}+r_{n}\right)^{2}}{4\Delta v}\right)\right]
\label{greensfunction}
\end{aligned}
\end{equation}
with
\begin{equation}
r_n=(-1)^nr+2nr_h, \quad \Delta v=\int_{E}^{E'}d\tilde{E} \frac{D(\tilde{E})}{b(\tilde{E})}.
\label{term2}
\end{equation}
The variable $\sqrt{\Delta v}$ has units of length and represents the mean distance traveled by an electron before losing energy of $\Delta E = E^{\prime} - E$. The $r_h$ denotes the diffusion-zone radius of the galaxy. The diffusion zone is a region where {CRs} propagate and has a size larger than that of the galaxy. The parameter $r_h$ depends on the spatial extent of both the gas and the magnetic field. However, there are currently uncertainties about the gas and magnetic properties of dSphs \cite{2015MNRAS.448.3747R}. In the Milky Way, the size of the diffuse region is several times larger than the width of the stellar disk. We assume that dSphs have a similar geometry and $r_h$ is given by $r_h = 2 r_{\rm max}$, where $r_{\rm max}$ is the distance from the dSph center to the outermost star. Previous work has indicated that the results are not greatly affected when $r_h$ varies by a factor of 0.5 to 2 \cite{Bhattacharjee_2021}. 
The parameters of the 8 classical dSphs we choose are listed in Table \ref{TableI}.

\begin{figure*}[htbp]
	\centering
	\includegraphics[width=0.45\textwidth]{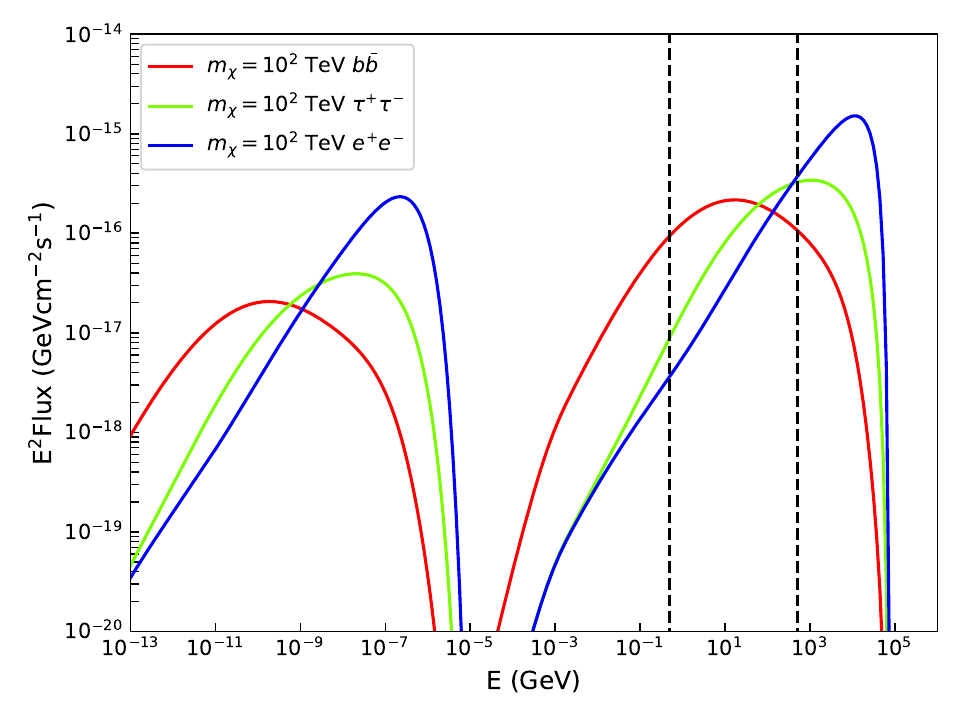}
	\includegraphics[width=0.45\textwidth]{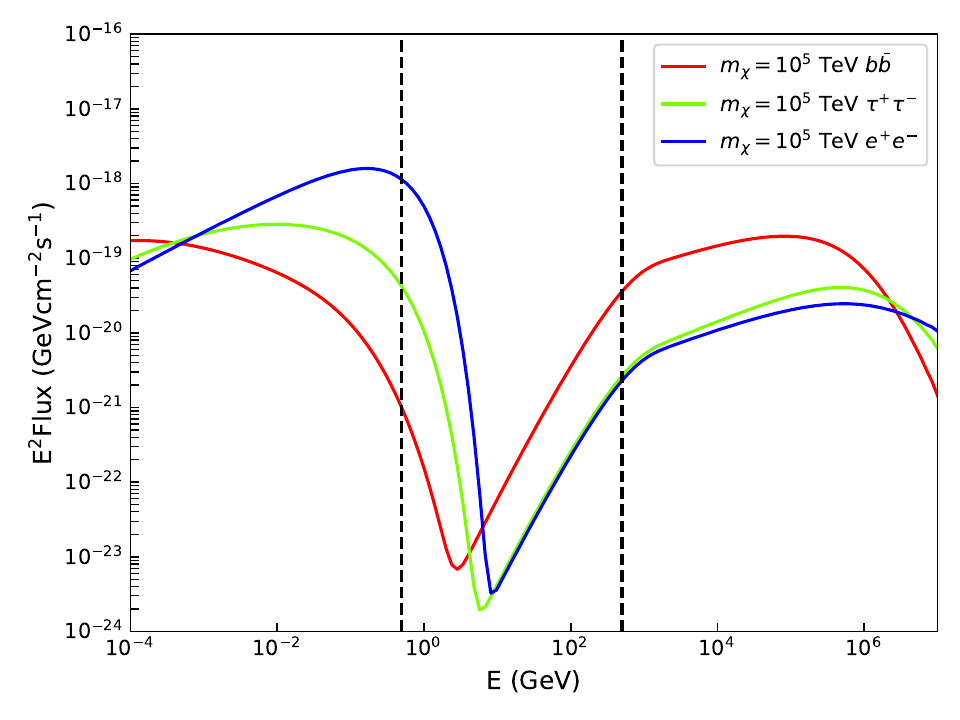}
	\caption{The SEDs of synchrotron and IC emission from the Draco dSph for two DM masses (left panel: $m_\chi$ = $10^2$ TeV, right panel: $m_\chi$ = $10^5$ TeV) and for the parameters of $B = 1 \,{\mu G}$ and $D_0 = 3\times10^{28}$ cm$^{2}$s$^{-1}$. Different color of lines represent three annihilation channels, $b\bar{b}$ (red line), $\tau^{+}\tau^{-}$ (green line) and $e^{+}e^{-}$ (blue line). The black dashed lines mark the Fermi-LAT energy band considered in this work (500 MeV - 500 GeV).}
	\label{fig:radiation}
\end{figure*}

\begin{figure*}[htbp]
	\centering
	\includegraphics[width=0.45\textwidth]{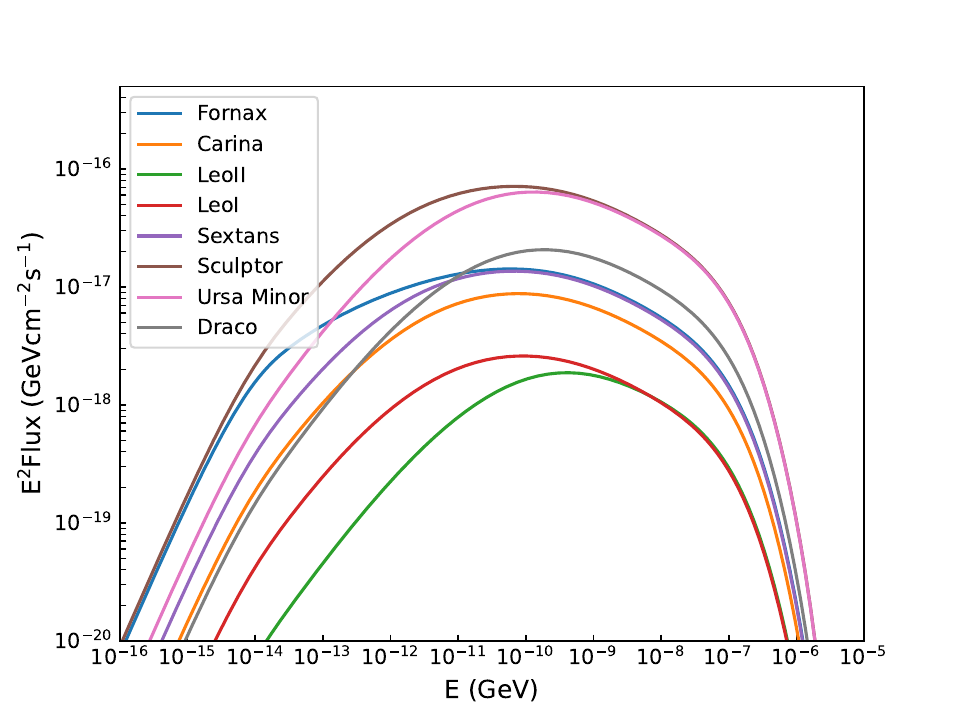}
        \includegraphics[width=0.45\textwidth]{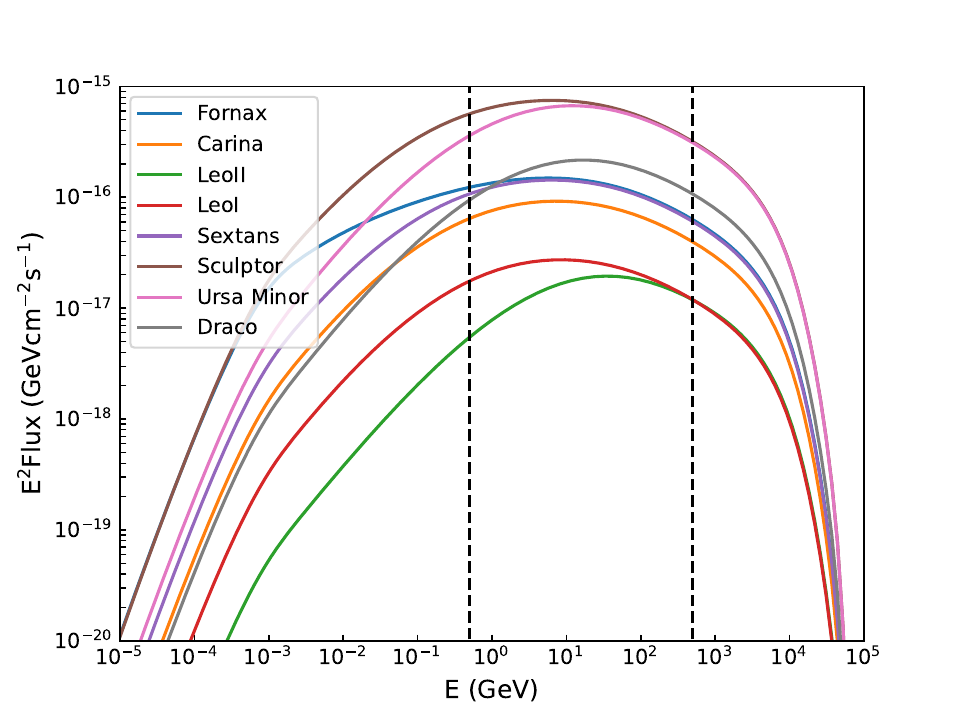}
	\caption{The SEDs of 8 sources for the $\chi\chi\rightarrow b\bar{b}$ channel and for $m_\chi=100$ TeV, $B = 1\,\mu$G, $D_0=3\times10^{28}\,\rm{cm^{2}s^{-1}}$ and $\langle \sigma v\rangle = 10^{-26}\,\rm{cm^{3}s^{-1}}$. The left panel shows the synchrotron SEDs, and the right panel shows the IC SEDs. The black dashed lines mark the energy band 500 MeV - 500 GeV.}
	\label{fig:8radiation}
\end{figure*}

For different particle energies, there are three different regimes for the solution of Eq.~(\ref{propagationequation}): no-diffusion, rapid-diffusion, and diffusion+cooling (see Appendix~\ref{app:a} for a comparison of the three solutions). In the no-diffusion regime, the electrons lose their energy much faster than they can diffuse in the dSph's magnetic field. In the rapid-diffusion scenario, the electrons diffuse very quickly in the magnetic field, so they can escape the galaxy without losing much energy. In the diffusion+cooling regime, both energy loss and diffusion are important processes, and they have similar time scales \cite{2006A&A...455...21C,2020PhRvD.101b3015K,Vollmann_2021}. In this work, we consider both the diffusion and cooling in our calculation. We first obtain the equilibrium distribution of electrons after diffusion and cooling using Eq.~(\ref{electrondnde}). The distributions integrated over the whole dSph for Draco are shown in \figurename{} \ref{fig:e_distribution}, where we consider three channels. We can see that the electron distribution of direct annihilation to $e^{+}e^{-}$ is higher than the other channels at high energies, but decreases at low energies. 

With the electron distribution in hand, we use the {\it Naima} package \cite{naima} to obtain the radiation spectrum.
The spectral energy distributions (SEDs) of the emission from Draco for $10^2$ TeV and $10^5$ TeV DM masses are shown in \figurename{} \ref{fig:radiation}, where we compare the spectra of three annihilation channels: $b\bar{b}$ (red line), $\tau^{+}\tau^{-}$ (green line) and $e^{+}e^{-}$ (blue line). The black dashed lines mark the energy band from 500 MeV to 500 GeV, namely the Fermi-LAT energy range considered in this work. The IC and synchrotron contributions dominate at high and low energies, respectively. From the SEDs of different DM masses, we can see that an $\mathcal{O}(100)$ TeV DM has a radiation peak within the Fermi-LAT energy band (especially when annihilating through $\chi\chi\rightarrow b\bar{b}$ or $\tau^{+}\tau^{-}$). The peak flux of the $e^{+}e^{-}$ channel is higher than the others. For $10^2$ TeV DM, the contribution is only from the IC emission. While as the mass increases to $\gtrsim10^5\,{\rm TeV}$, the IC peak is outside of the Fermi-LAT energy band and the contributions in the band are from both synchrotron and IC emission, with synchrotron becoming dominant.

The impact of the diffusion parameters and magnetic field on the SED of the secondary emission of the DM at the GeV-TeV scale has been widely discussed in {Refs. \cite{2020PhRvD.101b3015K,2023arXiv230314117W,2021JCAP...09..025C,2017JCAP...09..027M,2023JCAP...08..030R,2014JCAP...10..016R,2017JCAP...07..025R}.} The magnetic field distribution and cosmic-ray diffusion in dSphs are still poorly understood. {For the impact of the magnetic strength, a bigger magnetic field value results in a stronger signal strength for the synchrotron but has little effect on the IC component.
Unlike the magnetic strength affecting the synchrotron emission more than the IC emission, the diffusion coefficient $D_0$ affects both processes. A larger $D_0$ results in a weaker signal for both processes, because the relativistic charged particles can escape the diffusion region before losing much energy by synchrotron radiation and inverse Compton scattering.} For GeV scale DM, the peaks of the synchrotron radiation fall in the optical and radio energy bands, and for the works of using the radio observations to limit the DM one can see {Refs. \cite{2006A&A...455...21C,2007PhRvD..75b3513C,2017JCAP...09..027M,2023arXiv230314117W,2020PhRvD.101b3015K,2021JCAP...09..025C,PhysRevD.107.103011,2023JCAP...08..030R,2014JCAP...10..016R,2017JCAP...07..025R}.}

Different sources have different profile parameters ($\rho_s$, $r_s$), diffusion radii ($r_h$), and distances to the Earth, which affect the final constraints on the DM parameters. The SEDs of the 8 sources show differences because of these factors (\figurename{} \ref{fig:8radiation}). 
In \figurename{} \ref{fig:8radiation} we show the SEDs of the 8 dSphs for the annihilation through a $b\bar{b}$ channel with a cross section of $\left<\sigma v\right>= 10^{-26} \rm{cm}^{3}\rm{s}^{-1}$.

\section{Fermi-LAT data analysis}
We use 14 years (i.e. from 2008 August 21 to 2022 August 21) of Fermi-LAT data and select events from the Pass 8 SOURCE event class in the 500 MeV to 500 GeV energy range within a 10$^\circ$ radius of each dSph. The Fermi-LAT data is analyzed with the latest version of {\tt Fermitools} (Ver 2.2.0). To avoid the contamination of the Earth’s limb, events with zenith angles larger than 90$^\circ$ are rejected. The quality-filter cuts (DATA\_QUAL\textgreater0 \&\& LAT\_CONFIG==1) are applied to ensure the data can be used for scientific analysis. We take a 14$^\circ\times$14$^\circ$ square region of interest (ROI) for each target to perform a standard binned analysis. We consider all 4FGL-DR3 sources \cite{2020ApJS..247...33A} in ROI and two diffuse models (the Galactic diffuse gamma-ray emission {\tt gll\_iem\_v07.fits} and the isotropic component {\tt iso\_P8R3 SOURCE\_V3\_v1.txt}) to model the background. 

\begin{figure}
\centering
\includegraphics[width=0.98\columnwidth]{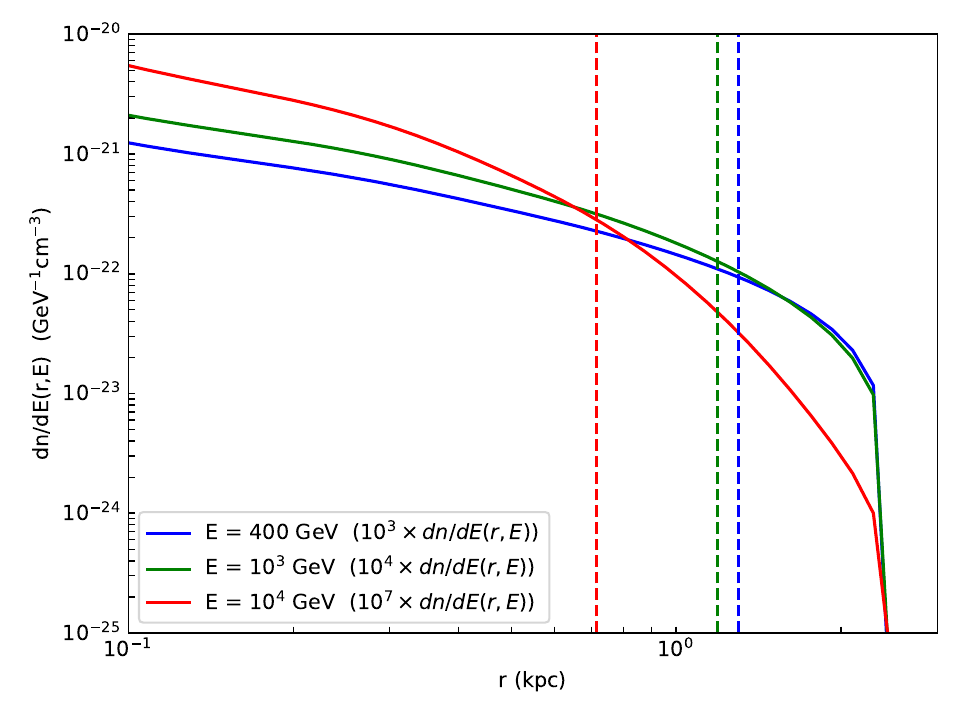}
\caption{The density distribution of the electrons from DM annihilation in Draco. Here we assume $m_\chi = 100\,\rm TeV$, $\langle\sigma v \rangle= 10 ^{-26}\,\rm cm^3 s^{-1}$ and $\chi\chi\rightarrow b\bar{b}$. 
The vertical lines mark the $68\%$ containment radii of the electron distribution, corresponding to angular radii of $0.54^\circ$ (red dashed), $0.9^\circ$ (green dashed), $1.0^\circ$ (blue dashed) in the sky.} 
\label{dndErE}
\end{figure}

\begin{figure}
\centering
\includegraphics[width=0.98\columnwidth]{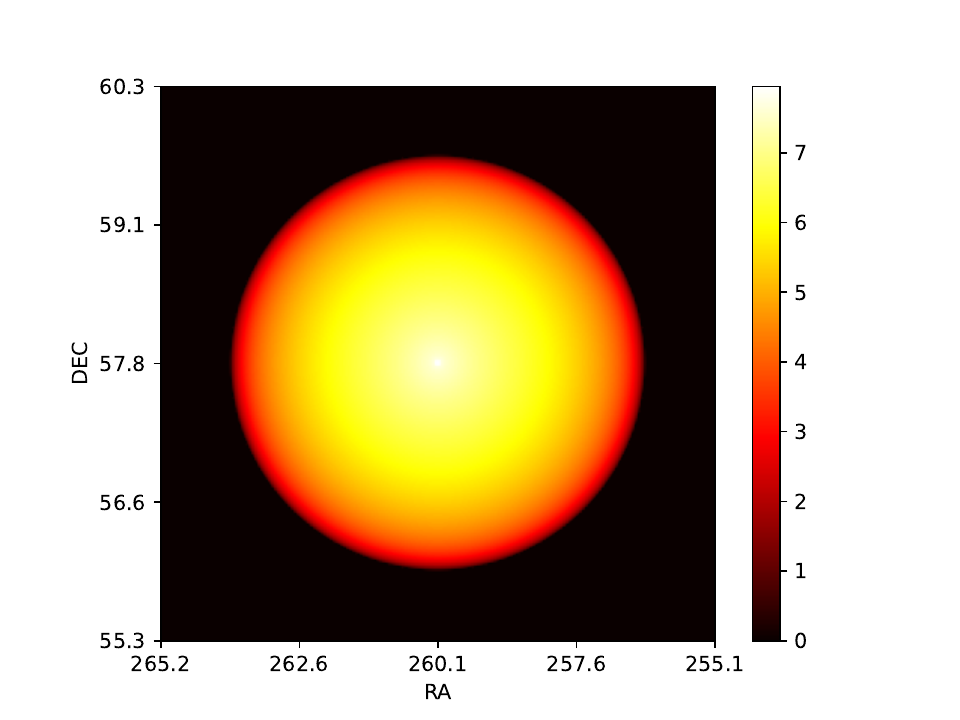}
	\caption{The normalized spatial template at 500 MeV used in the Fermi-LAT data analysis for the Draco dSph.}
	\label{template}
\end{figure}

{Due to the diffusion of CR electrons in dSphs, the emission of gamma rays presents as extended sources in the sky.
As an illustration, \figurename{} \ref{dndErE} shows that the spatial distribution of electrons in Draco is more extended than the Fermi-LAT PSF.
The 68\% containment angles of the extended gamma-ray emissions at 500 MeV for the 8 dSphs under benchmark parameters ($D_0=3\times10^{28}\,\rm cm^2\,s^{-1}$, $B=1\,\rm \mu G$) are listed in Table~\ref{TableI}, which are all larger than the Fermi-LAT angular resolution.
For this reason, we model dSphs as extended sources in the Fermi-LAT data analysis.
We create the spatial templates used in the Fermi-LAT data analysis based on the expected gamma-ray fluxes at different directions around the dSphs.
\figurename{} \ref{template} demonstrates the template at 500 MeV for the Draco dSph.
}

\begin{figure*}[htbp]
	\centering
        \includegraphics[width=0.6\textwidth]{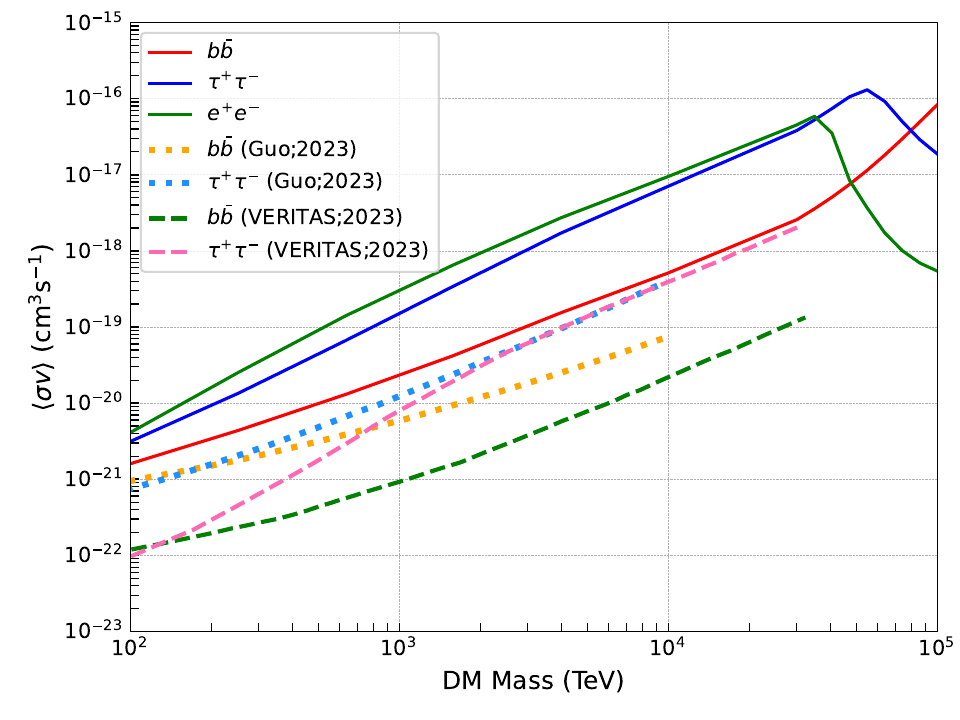}
	\caption{Constraints on the DM annihilation cross section at the 95\% CL for the $b\bar{b}$ (red line), $\tau^{+}\tau^{-}$ (blue line) and $e^{+}e^{-}$ (green line) channels. These constraints are derived from a combined analysis of 8 dSphs based on the null detection of the secondary IC and synchrotron emission of DM annihilation from the dSphs. Also plotted for comparisons are results from Refs.~\cite{2023ApJ...945..101A,PhysRevD.108.043001}.}
	\label{fig:diff_result}
\end{figure*}

We perform a binned Poisson maximum likelihood analysis in 24 logarithmically-spaced bins of energy from 500 MeV to 500 GeV, with a spatial pixel size of 0.1$^\circ$. The likelihood function for the $i$th target is given by
\begin{equation}
\mathcal{L}_{i}=\prod_{k}\frac{\lambda_k^{n_{k}}e^{-\lambda_k}}{n_k!},
\label{eq:like}
\end{equation}
where  
$\lambda_k$ is the model-predicted photon counts and $n_k$ is the observational photon counts with $k$ the index of the energy and spatial bins. 
The model-predicted photon counts $\lambda_k$ incorporates both the contributions from the background and the (possible) DM signal, which for the $k$th bin is given by
\begin{equation}
\lambda_k = n_1N^{\mathrm{bkg}}_{k}+n_2N^{\mathrm{DM}}_{k}
\end{equation}
where $N^{\rm{bkg}}_{k}$ is the photon counts from the background (including 4FGL sources and two diffuse backgrounds) and $N^{\mathrm{DM}}_{k}$ is the photon counts of the DM signal. The $n_1$ is the rescaling factor of the background component, which is introduced to account for possible systematic uncertainties in the best-fit background model. The $n_2$ is the free parameter of the DM component. The $N^{\rm{bkg}}_{k}$ and $N^{\rm{DM}}_{k}$ are obtained using the {\tt gtmodel} command in the {\tt Fermitools} software.The DM component is implemented in the analysis using a {\tt FileFunction} spectrum, while the required DM spectral files are obtained via {\tt HDMSpectra}.

We first utilize the standard Fermi-LAT binned likelihood analysis\footnote{\url{https://fermi.gsfc.nasa.gov/ssc/data/analysis/scitools/binned_likelihood_tutorial.html}} to obtain the best-fit background model (with no DM component added). During the fitting procedure, we free the parameters of all 4FGL-DR3 \cite{2020ApJS..247...33A} sources within a $14^\circ\times14^\circ$ ROI, as well as the normalizations of the two diffuse components, and use the {\tt NewMinuit} optimizer to perform the fitting. After obtaining the best-fit background model, for a given DM mass $m_\chi$, we use {\tt gtmodel} to generate $N^{\rm{bkg}}_{k}$ and $N^{\rm{DM}}_{k}$. We scan a series of values of $n_2$, and for each $n_2$ we fit the $n_1$ to maximize the likelihood in Eq.~(\ref{eq:like}), obtaining the change to the likelihood as a function of the $n_2$ (likelihood profile). Varying the DM mass and repeating this process, we obtain ${\mathcal{L}}(n_2)$ for different DM masses. We finally obtain a Likelihood grid that is related to a range of $m_\chi$ and $\left<\sigma v\right>$ values ($n_2\propto\left<\sigma v\right>$), covering all DM parameters in the analysis.

Based on this likelihood grid we can determine the best-fit $m_\chi$ and $\left<\sigma v\right>$, compute the test statistic (TS) value of the target, as well as set upper limits on the $\left<\sigma v\right>$. The likelihood grid will also be used for the subsequent combined analysis. The TS is defined as $\rm{TS} = -2\ln(\mathcal{L}_{\rm{bkg}} - \mathcal{L}_{\rm{dsph}})$ \cite{1996ApJ...461..396M}, where $\mathcal{L}_{\rm{bkg}}$ and $\mathcal{L}_{\rm{dsph}}$ are the best-fit likelihood values for the background-only model and the model with a dSph included, respectively. The upper limit on $\left<\sigma v\right>$ for a given DM mass corresponds to the value of $\left<\sigma v\right>$ that makes the best-fit $-\ln\mathcal{L}$ increased by 1.35.

The combined analysis can improve the sensitivity of the analysis by staking sources, it assumes that the properties of DM particles are identical for all dSphs \cite{2011PhRvL.107x1302A,2014PhRvD..89d2001A, Fermi2015prl}. 
The combined likelihood function is
\begin{equation}
\widetilde{\mathcal{L}}(\langle\sigma v\rangle,m_{\chi}) = \prod_{i}\mathcal{L}_{i}(\langle\sigma v\rangle,m_{\chi})
\label{eq:clike}
\end{equation}
with $\mathcal{L}_{i}$ the likelihood in Eq.~(\ref{eq:like}) for the $i$th source.
For the analysis of the prompt $\gamma$-rays of DM annihilation (rather than the secondary IC and synchrotron emission), we also consider the uncertainties on the J-factors by including an additional term in the likelihood function, namely \cite{2014PhRvD..89d2001A,Fermi2015prl}
\begin{equation}
	\mathcal{L} = \prod_{i}\mathcal{L}_{i}(\langle\sigma v\rangle,m_{\chi}) \times \mathcal{L}_{J}(J_{i}|J_{\mathrm{obs},i},\sigma_{i})
\end{equation}
and
\begin{equation}
	\mathcal{L}_{J}(J_{i}|J_{\mathrm{obs},i},\sigma_{i})  = \frac{1}{\ln(10)J_{\mathrm{obs},j}\sqrt{2\pi}\sigma_{i}} \times  {e}^{-\frac{(\log_{10}J_{i}-\log_{10}J_{{\rm obs},i})^{2}}{2\sigma_{i}^{2}}}
\end{equation}
where $J_{\mathrm{obs},i}$ is the measured J-factor with its uncertainty $\sigma_{i}$ and $J_{i}$ is the true value of the J-factor which is to be determined in the fitting. One can see Refs.~\cite{2014PhRvD..89d2001A,Fermi2015prl} for more details of the combined likelihood analysis.

\section{results and discussion}
\label{sec:res}
We find no significant (i.e. TS$\geq$25) gamma-ray signal from the secondary emission of DM annihilation in the directions of the 8 dSphs. For a series of DM masses of $10^2 - 10^5$ TeV, we derive the 95\% CL upper limits for the DM annihilation cross section of $e^+e^-$, $b\bar{b}$ and $\tau^+\tau^-$ channels. For the single-source analysis, we find that the observations of Sextans and Fornax give the strongest constraints for the $b\bar{b}\,/\,\tau^+\tau^-$ and $e^+e^-$ channels, respectively.
The combined analysis of eight dSphs yields constraints about three times stronger than the best single-source constraints. The results of the combined analysis are shown in \figurename{} \ref{fig:diff_result}. For the $e^+e^-$ or $\tau^+\tau^-$ channel, the exclusion line in the figure has a break close to $m_\chi=10^5\,{\rm TeV}$, mainly because the synchrotron radiation starts to become dominant over IC as the mass increases to $\gtrsim3\times10^4\,{\rm TeV}$.

{Compared to the existing limits in the $10^2-10^5\,{\rm TeV}$ mass range, our constraints are weaker than the ones obtained from the VERITAS and IceCube observations of dSphs \cite{2023ApJ...945..101A,PhysRevD.108.043001}. However, our analysis can extend the constraints on HDM to the larger mass range.}

\begin{figure*}
	\centering
	\includegraphics[width=0.39\textwidth]{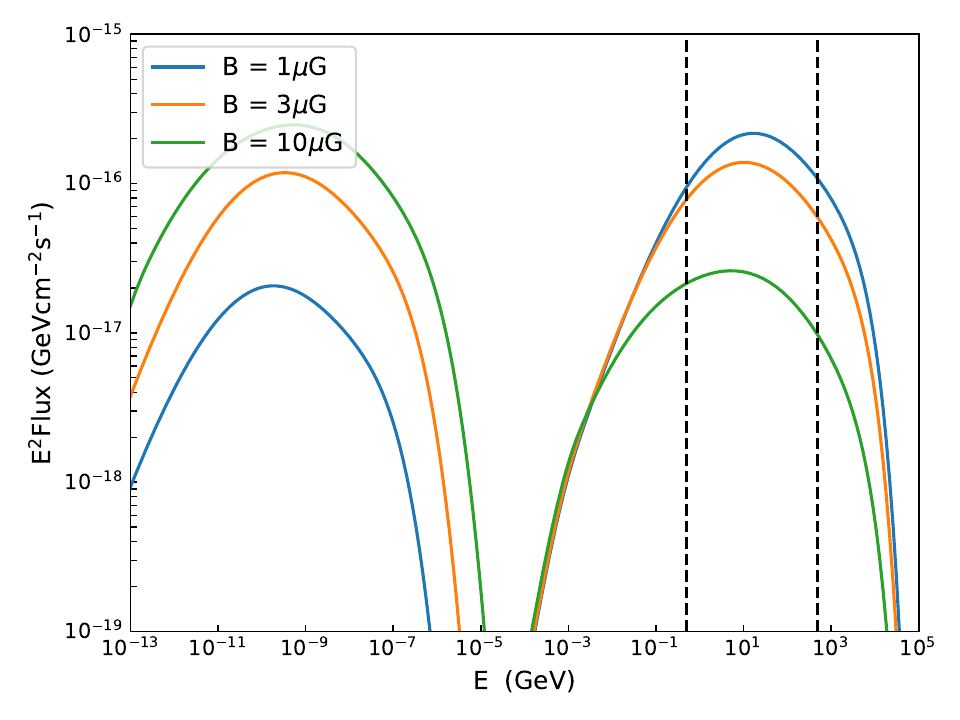}
	\includegraphics[width=0.39\textwidth]{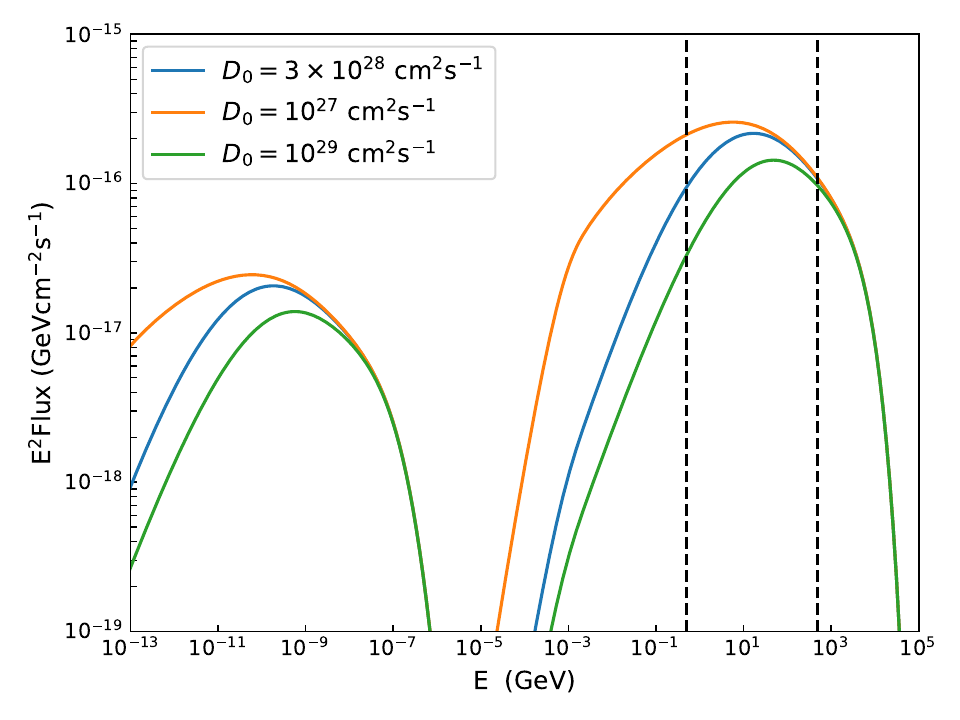}\\
    \includegraphics[width=0.39\textwidth]{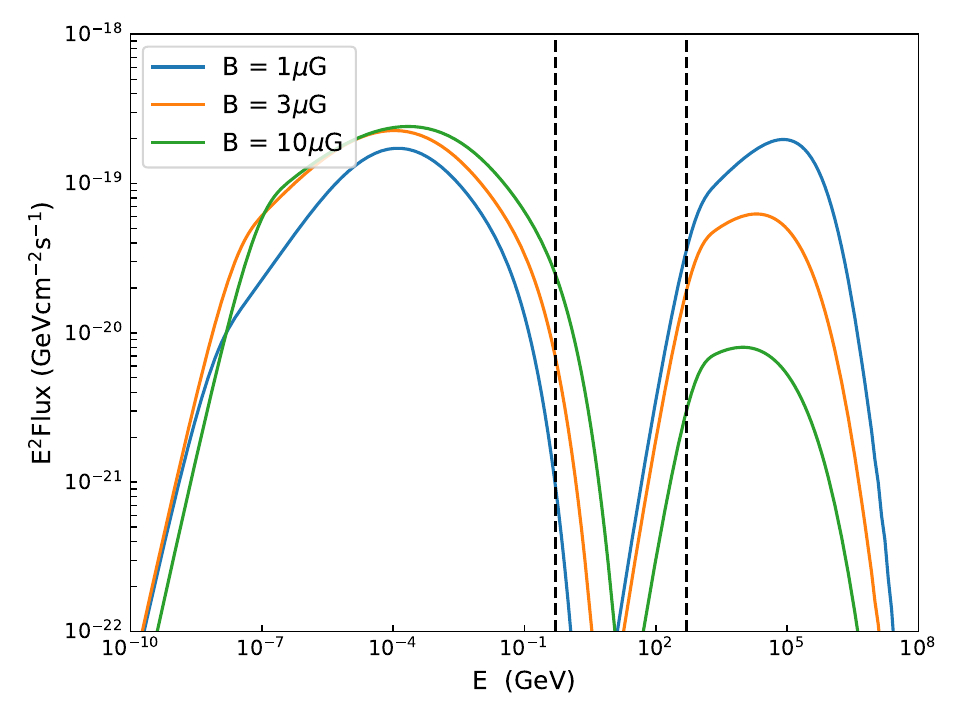}
	\includegraphics[width=0.39\textwidth]{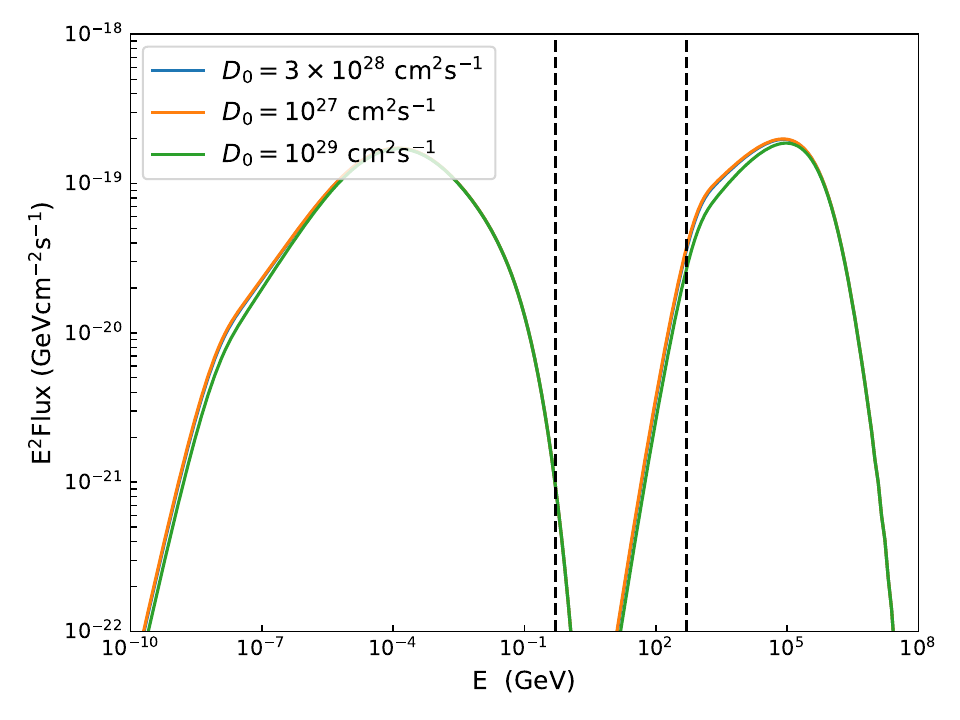}
	\caption{We compare the model SEDs for different values of magnetic field (left panels, $D_0 =  3 \times 10^{28}\,\rm{cm^2\,s^{-1}}$) and diffusion coefficient (right panels, $B = 1\,\mu \rm G$) for the Draco dSph. We assume three values of magnetic field and diffusion coefficient for the $b \bar{b}$ channel. The upper and lower panels are for $m_\chi = 100 \rm \ TeV$ and $m_\chi = 100 \rm \ PeV$, respectively. The cross section is $\langle\sigma v \rangle= 10 ^{-26} \rm \ cm^3\,s^{-1}$. The black dashed lines mark the energy band 500 MeV $-$ 500 GeV.}
	\label{BD1}
\end{figure*}

It should be noted that the constraints on DM parameters via the secondary emission of DM annihilation are affected by the uncertainties of model parameters. Possible sources of uncertainty include the diffusion coefficient ($D_0$), the magnetic field strength ($B$), and the size of the diffusion-zone ($r_h$). The value of $D_0$ is typically in the range of $10^{27}-10^{29}\,{\rm cm^2s^{-1}}$ \cite{2020A&A...639A.131W,2019PhLB..789..292W,2019PhRvD..99j3023E,1971ApJ...163..255P,2016ApJ...824...16J}, while the possible range of $B$ is $1-10\,{\rm \mu G}$ \cite{2021JCAP...09..025C,2015MNRAS.448.3747R}. 
{In deriving the results of FIG.~\ref{fig:diff_result}, we have made specific assumptions about the magnetic field and
diffusion strength ($D_0=3\times10^{28}\,\rm cm^2\,s^{-1}$, $B=1\,\rm \mu G$).
To demonstrate the influence of $B$ and $D_0$ on the results, we present Fig.~\ref{BD1}, in which we show how the model-expected SED will change when choosing different $B$ and $D_0$ values.

Since the strength of the magnetic field affects both the synchrotron emissivity of a single electron, as well as the energy loss rate of the electron during propagation, it can be seen that the value of $B$ in dSphs will have a large effect on the results.
If dSphs have a larger $B$ value, our constraints will be weaker. For the parameter $D_0$, the electrons produced by the annihilation of HDM are fairly energetic and lose most of their energy before they can diffuse efficiently. Therefore, the effect of the diffusion coefficient is relatively minor, as shown in the right panels of FIG.~\ref{BD1} (see also Appendix~\ref{app:b}). Only at the low mass end of the mass range we are considering (several hundred TeV), the value of the diffusion coefficient will have a certain influence on the results.}

\begin{table}[h]
	\centering{}
	\caption{The sample of 15 Milky Way dSphs that used for the analysis of prompt gamma rays of DM annihilation. The J-factors are extracted from Ref.~\cite{Fermi2015prl}.}
	\begin{ruledtabular}
		\begin{tabular}{lcccc}
			Name & \multicolumn{1}{c}{\begin{tabular}[c]{@{}c@{}}$l$\\ (deg)\end{tabular}} & \multicolumn{1}{c}{\begin{tabular}[c]{@{}c@{}}$b$\\ (deg)\end{tabular}} & \multicolumn{1}{c}{\begin{tabular}[c]{@{}c@{}}Distance\\ (kpc)\end{tabular}} & \multicolumn{1}{c}{\begin{tabular}[c]{@{}c@{}} ${\log}_{10}(J_{\mathrm{obs}})$ \\ ($\rm{log}_{10}$[$\rm{GeV}^2$ $\rm{cm}^{-5}]$)\end{tabular}} \\
			\colrule
			Bootes I & 358.1 & 69.6 & 66 & $18.8 \pm 0.22$ \\
			Canes Venatici II & 113.6 & 82.7 & 160 & $17.9 \pm 0.25$ \\
			Carina & 260.1 & $-22.2$ & 105 & $18.1 \pm 0.23$ \\
			Coma Berenices & 241.9 & 83.6 & 44 & $19.0 \pm 0.25$ \\
			Draco & 86.4 & 34.7 & 76 & $18.8 \pm 0.16$ \\
			Fornax & 237.1 & $-65.7$ & 147 & $18.2 \pm 0.21$ \\
			Hercules & 28.7 & 36.9 & 132 & $18.1 \pm 0.25$ \\
			Leo II & 220.2 & 67.2 & 233 & $17.6 \pm 0.18$ \\
			Leo IV & 265.4 & 56.5 & 154 & $17.9 \pm 0.28$ \\
			Sculptor & 287.5 & $-83.2$ & 86 & $18.6 \pm 0.18$ \\
			Segue 1 & 220.5 & 50.4 & 23 & $19.5 \pm 0.29$ \\
			Sextans & 243.5 & 42.3 & 86 & $18.4 \pm 0.27$ \\
			Ursa Major II & 152.5 & 37.4 & 32 & $19.3 \pm 0.28$ \\
			Ursa Minor & 105.0 & 44.8 & 76 & $18.8 \pm 0.19$ \\
			Willman 1 & 158.6 & 56.8 & 38 & $19.1 \pm 0.31$ \\
		\end{tabular}
	\end{ruledtabular}
	\label{TableII}
\end{table}
 
Other sources of uncertainty include the accuracy of the modeling of the DM halos. For example, this paper considers a spherical NFW profile; if it were not perfectly spherical, the results would vary by about tens of percent \cite{PhysRevD.94.063521}.

\begin{figure*}[t]
\centering
\includegraphics[width=0.45\textwidth]{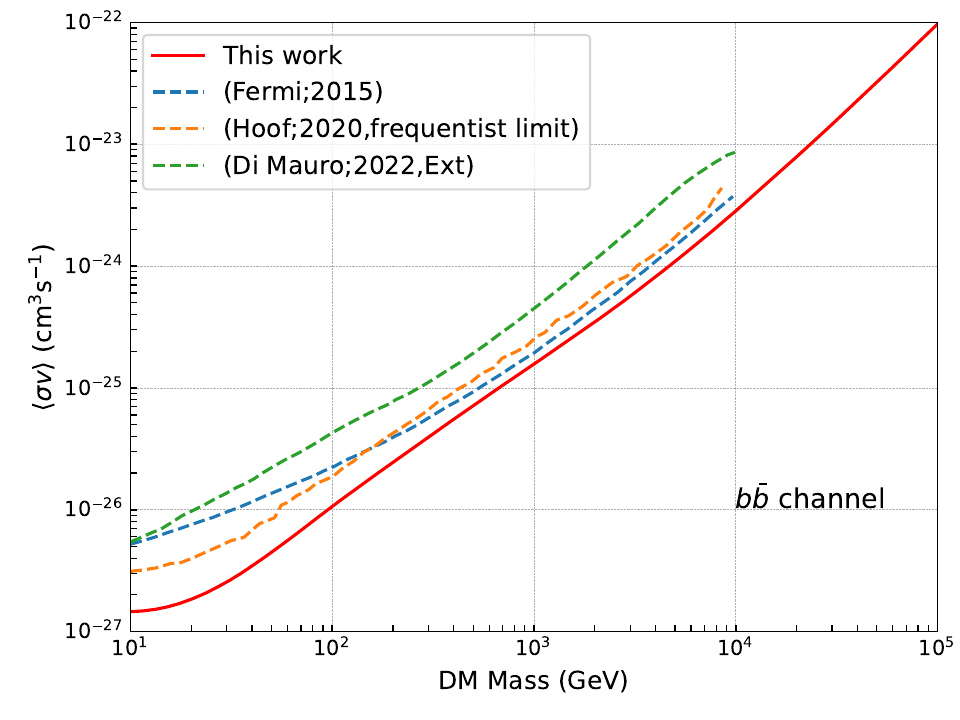}
\includegraphics[width=0.45\textwidth]{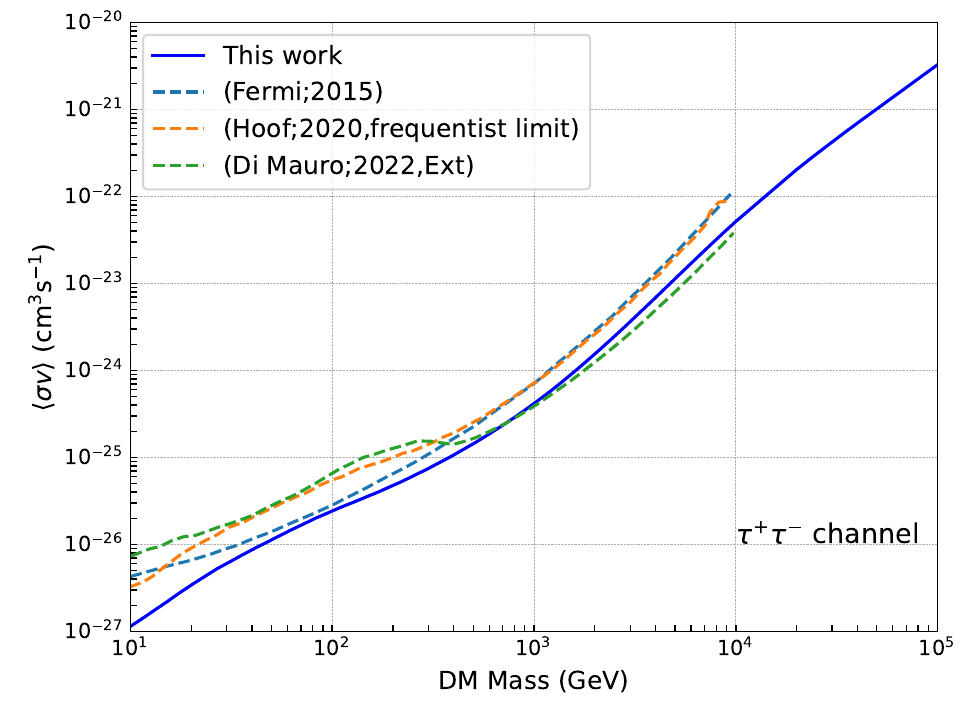}
\caption{Constraints on the DM annihilation cross section at the 95\% CL for the $b\bar{b}$ (left panel) and $\tau^{+}\tau^{-}$ (right panel) channels. These constraints are derived from a combined analysis of 15 dSphs based on the null detection of the prompt gamma rays of DM annihilation from the dSphs. Also plotted for comparisons are results from Refs.~\cite{Fermi2015prl,2020JCAP...02..012H,PhysRevD.106.123032}. Note that the results of Ref.~\cite{PhysRevD.106.123032} consider the extension of dSphs, which gives weaker (by a factor of $\sim$2) but more realistic constraints on DM parameters. {\it Caution: The units of the x-axis in this figure are different from FIG.~\ref{fig:diff_result}.}}
\label{fig:result_15}
\end{figure*}

{Lastly, as a complement, we also search for the prompt $\gamma$-ray signals (i.e., gamma-ray photons are produced by the hadronization or decay of final-state particles, not by secondary synchrotron or IC emission of electrons) of DM annihilation from a larger sample of 15 dSphs and set constraints on the cross section for DM in the GeV-TeV mass range.
The samples are the same as that used in Ref.~\cite{Fermi2015prl} and are listed in Table \ref{TableII}.
For this part of the analysis that is concerned with the prompt signals, we model all dSphs as point-like sources to ease of comparison with most of the previous results in the literature. 
Note however that Ref.~\cite{PhysRevD.106.123032} pointed out the limits on DM parameters will be weakened by a factor of $\sim1.5-2.5$ if modeling the dSphs as extended sources.}

For the case of prompt $\gamma$-rays, the expected $\gamma$-ray flux from DM annihilation can be written as
\begin{equation}
\Phi=\underbrace{\frac{1}{4\pi}\frac{\langle\sigma v\rangle}{2 m_{\chi}^{2}}\frac{d N_{\gamma}}{d E_{\gamma}}}_{\rm particle\ physics}\times\underbrace{\int_{\Delta\Omega}\int_{\rm LOS}\rho^{2}(r)dld{\Omega}}_{\rm J-factor}
\label{dmflux}
\end{equation}
where the first particle physics term depends on the DM mass $m_{\chi}$, the thermally-averaged annihilation cross section $\langle\sigma v\rangle$ and the differential $\gamma$-ray yield per annihilation ${dN_{\gamma}}/{dE_{\gamma}}$. We use {\tt PPPC4DMID} \cite{2011JCAP...03..051C} to generate spectra for GeV-TeV DM annihilation. The second term is related only to the astrophysical distribution of DM (called J-factor), which is the line of sight integral of the squared DM density $\rho(r)^{2}$ over a solid angle ${\Omega}$. The J-factors can be derived from stellar kinematic data. The accurate determination of J-factors is crucial for using the observations of dSphs to study the DM properties \cite{2013NewAR..57...52B, 2008ApJ...675..201M, 2009ApJ...704.1274W}. 
In this work, we adopt the same J-factor values as in \cite{Fermi2015prl} for the 15 dSphs. The parameters of the dSphs are listed in Table \ref{TableII}.

For the analysis of individual sources, the two dSphs, Segue 1 and Ursa Major II, provide the strongest constraints, while the combined analysis of the 15 dSphs gives even stronger limits. \figurename{} \ref{fig:result_15} shows the results of the combined analysis for DM annihilation through the $b\bar{b}$ and $\tau^+\tau^-$ channels from $m_\chi=10\,{\rm GeV}$ to $10^5$ GeV. Our results are stronger than the existing limits in the literature \cite{Fermi2015prl}. The improvement is due to the use of a larger data set as well as an updated version of the Fermi-LAT data (Pass 8), instrument response function (P8R3\_V3), and diffuse background models.

{\it A short summary.}
In this paper, based on the up-to-date Fermi-LAT observations of dSphs, constraints on the DM annihilation cross section are derived for a very wide mass range from 10 GeV to 100 PeV in a unified framework of Fermi-LAT data analysis. Among the whole mass range, the results of the $10^2-10^5$ GeV part (\figurename{}~\ref{fig:result_15}) are obtained by considering the gamma-ray signals directly produced by dark matter annihilation, while the $100-10^5$ TeV mass range (\figurename{}~\ref{fig:diff_result}), which is the main focus of this paper, takes into account the secondary radiation produced by heavy dark matter through the inverse-Compton and synchrotron processes.
Although the constraints we obtain are weaker than the previous limits given by the VERITAS and IceCube observations of dSphs, we extend the existing limits to higher DM masses.
We also discuss the influence of the choice of the values of $B$ and $D_0$ on the results.

{We are aware of a similar work that appeared online \cite{song2024search} when we were preparing the response to the referee.
Note that, following the referee's suggestion, in this version we also model the dSphs as extended sources when performing the analysis of the secondary emission. 
We notice that the two works get similar constraints on the DM parameters.}

\begin{acknowledgments}
        We acknowledge data and scientific data analysis software provided by the Fermi Science Support Center.
	This work is supported by the National Key Research and Development Program of China (No. 2022YFF0503304) and the Guangxi Science Foundation (grant No. 2019AC20334).
\end{acknowledgments}

\bibliographystyle{apsrev4-1-lyf}
\bibliography{Reference.bib}

\newpage
\widetext
\appendix
\renewcommand{\figurename}{FIG.}
\renewcommand{\thefigure}{S\arabic{figure}}
\setcounter{figure}{0}
\section{Solutions for the no-diffusion and rapid-diffusion approximations}
\label{app:a}
For the no-diffusion scenario, the solution is \cite{2022PhRvD.105d3011Z,2023PhRvD.107l3027Z}
$$\frac{d n}{d E}(r,E)=\frac{1}{b(E)}\int_{E}^{m_{\chi}}d E^{\prime}Q_{e}(E^{\prime},r),$$
where $b(E)$ is the energy loss rate and $Q_{e}(E^{\prime},r)$ is the electron injection.
For the rapid-diffusion scenario, the solution is \cite{Vollmann_2021} $$\frac{d n}{d E}(r,E)=\frac{1}{D(E)}\int\mathrm{d}r^{\prime}\frac{r^{\prime}}{r}\left(\frac{1}{2}(r+r^{\prime})-\frac{1}{2}|r-r^{\prime}|-\frac{r r^{\prime}}{r_{h}}\right)Q_{e}(E,r^{\prime}),$$
where $r_h$ is the size of the diffusion zone and $D(E)$ is the diffusion coefficient.

\section{Time scales for different processes in CR propagation}
\label{app:b}

\begin{figure}[h]
\centering
\includegraphics[width=0.49\textwidth]{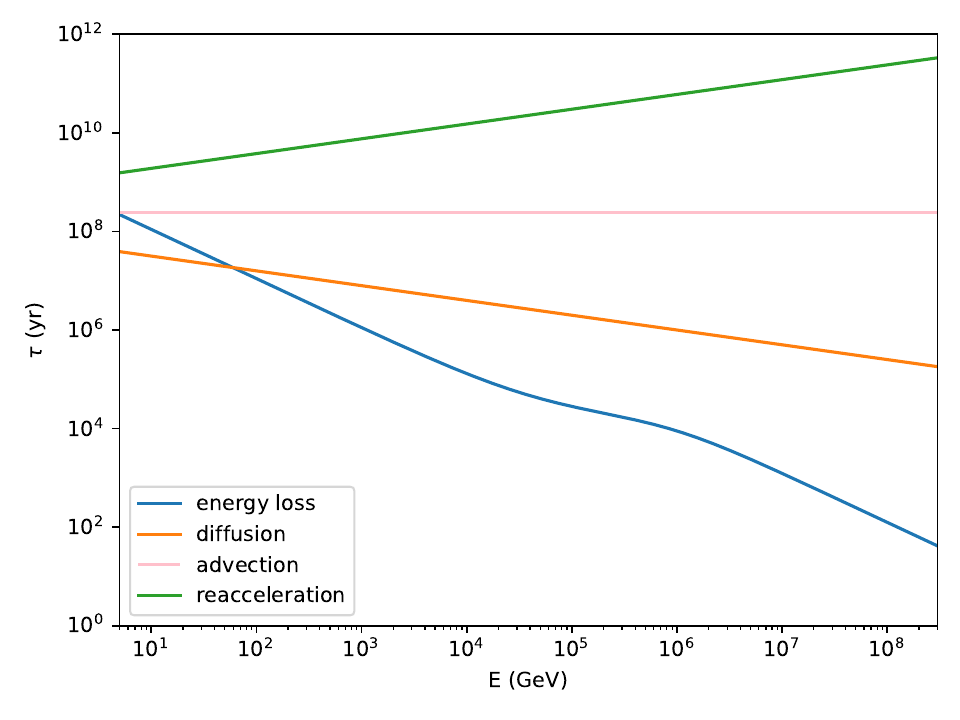}
\caption{The time scales for different processes: energy loss (blue), diffusion (orange), advection (pink), reacceleration (green).}
\label{time}
\end{figure}

In Fig.~\ref{time}, we calculate the time scales for different processes related to the propagation of CR electrons using the following equations \cite{2023JCAP...08..030R,2016PASJ...68...34F,2022ApJ...941...65B},
\begin{equation}
\tau_{\rm{loss}} \approx \frac{E}{b(E)},\ 
\tau_{\rm{diff}} \approx \frac{r_h^2}{D(E)},\ 
\tau_{\rm {acc}} \approx  \frac{p^2}{D_{pp}},\ 
\tau_{\rm {adv}} \approx \frac{r_{h}^2}{ r_h v_{a}/3}.
\end{equation}
The $\tau_{\rm{loss}}$, $\tau_{\rm{diff}}$, $\tau_{\rm {acc}}$ and $\tau_{\rm {adv}}$ are the time scales for energy loss, spatial diffusion, reacceleration, and advection, respectively.
Here, we assume $D_{0} = 3 \times 10^{28}\, \rm{cm/s}$, $r_h=2.5\,\rm kpc$, and $v_{a} = 30\, \rm{km/s}$. 
From Fig.~\ref{time}, it can be seen that the advection and reacceleration are subdominant compared to energy loss and spatial diffusion at the energies we are interested in and therefore can be ignored in our calculation.

\end{document}